\documentclass[twocolumn,tighten]{aastex62}

\shorttitle{Two flares with one shock in 3C~454.3}
\shortauthors{Liodakis et al.}
\begin{document}

\title{Two Flares with One Shock: the Interesting Case of 3C~454.3}

\correspondingauthor{I. Liodakis}
\email{ilioda@stanford.edu}

\author[0000-0001-9200-4006]{I. Liodakis}
\affil{KIPAC, Stanford University, 452 Lomita Mall, Stanford, CA 94305, USA}

\author[0000-0003-0611-5784]{D. Blinov}
\affil{Institute of Astrophysics, Foundation for Research and Technology-Hellas, GR-71110 Heraklion,Greece}
\affil{Department of Physics, Univ. of Crete, GR-70013 Heraklion, Greece}
\affil{Astronomical Institute, St.  Petersburg State University, Universitetskij Pr.  28, Petrodvorets, 198504 St.  Petersburg, Russia}

\author{S. G. Jorstad}
\affil{Astronomical Institute, St.  Petersburg State University, Universitetskij Pr.  28, Petrodvorets, 198504 St.  Petersburg, Russia}
\affil{Institute for Astrophysical Research, Boston University, 725 Commonwealth Avenue, Boston, MA02215}

\author{A.~A.~Arkharov}
\affil{Pulkovo Observatory, St.-Petersburg, Russia}

\author{A.~Di Paola}
\affil{INAF, Osservatorio Astronomico di Roma, Italy}

\author{N.~V.~Efimova}
\affil{Pulkovo Observatory, St.-Petersburg, Russia}

\author[0000-0002-3953-6676]{T.~S.~Grishina}
\affil{Astronomical Institute, St.  Petersburg State University, Universitetskij Pr.  28, Petrodvorets, 198504 St.  Petersburg, Russia}

\author{S. Kiehlmann}
\affil{Institute of Astrophysics, Foundation for Research and Technology-Hellas, GR-71110 Heraklion,Greece}
\affil{Department of Physics, Univ. of Crete, GR-70013 Heraklion, Greece}

\author[0000-0001-9518-337X]{E.~N.~Kopatskaya}
\affil{Astronomical Institute, St.  Petersburg State University, Universitetskij Pr.  28, Petrodvorets, 198504 St.  Petersburg, Russia}

\author[0000-0002-4640-4356]{V.~M.~Larionov}
\affil{Astronomical Institute, St.  Petersburg State University, Universitetskij Pr.  28, Petrodvorets, 198504 St.  Petersburg, Russia}
\affil{Pulkovo Observatory, St.-Petersburg, Russia}

\author[0000-0002-0274-1481]{L.~V.~Larionova}
\affil{Astronomical Institute, St.  Petersburg State University, Universitetskij Pr.  28, Petrodvorets, 198504 St.  Petersburg, Russia}

\author[0000-0002-2471-6500]{E.~G.~Larionova}
\affil{Astronomical Institute, St.  Petersburg State University, Universitetskij Pr.  28, Petrodvorets, 198504 St.  Petersburg, Russia}

\author{A.~P. Marscher}
\affil{Institute for Astrophysical Research, Boston University, 725 Commonwealth Avenue, Boston, MA02215}

\author[0000-0001-9858-4355]{D.~A.~Morozova}
\affil{Astronomical Institute, St.  Petersburg State University, Universitetskij Pr.  28, Petrodvorets,  198504 St.  Petersburg, Russia}

\author[0000-0001-9858-4355]{A.~A.~Nikiforova}
\affil{Astronomical Institute, St.  Petersburg State University, Universitetskij Pr.  28, Petrodvorets, 198504 St.  Petersburg, Russia}
\affil{Pulkovo Observatory, St.-Petersburg, Russia}

\author{V. Pavlidou}
\affil{Institute of Astrophysics, Foundation for Research and Technology-Hellas, GR-71110 Heraklion,Greece}
\affil{Department of Physics, Univ. of Crete, GR-70013 Heraklion, Greece}

\author{E. Traianou}
\affil{Max-Planck-Institut f\"ur Radioastronomie, Auf dem H\"ugel 69, D-53121, Bonn, Germany}

\author[0000-0002-9907-9876]{Yu.~V.~Troitskaya}
\affil{Astronomical Institute, St.  Petersburg State University, Universitetskij Pr.  28, Petrodvorets, 198504 St.  Petersburg, Russia}

\author[0000-0002-4218-0148]{I.~S.~Troitsky}
\affil{Astronomical Institute, St.  Petersburg State University, Universitetskij Pr.  28, Petrodvorets, 198504 St.  Petersburg, Russia}

\author{M. Uemura}
\affil{Hiroshima Astrophysical Science Center, Hiroshima University, 1-3-1 Kagamiama,Higashi-Hiroshima, 739-8526, Japan}

\author[0000-0001-6314-0690]{Z.~R. Weaver}
\affil{Institute for Astrophysical Research, Boston University, 725 Commonwealth Avenue, Boston, MA02215}

\begin{abstract}
The quasar 3C~454.3 is a blazar known for its rapid and violent outbursts seen across the electromagnetic spectrum. Using $\gamma$-ray, X-ray, multi-band optical, and very long baseline interferometric data we investigate the nature of two such events that occurred in 2013 and 2014 accompanied by strong variations in optical polarization, including a $\rm\sim230^o$ electric vector position angle (EVPA) rotation. Our results suggest that a single disturbance was responsible for both flaring events. We interpret the disturbance as a shock propagating down the jet. Under this interpretation the 2013-flare originated most likely due to changes in the viewing angle caused by perhaps a bent or helical trajectory of the shock upstream of the radio core.  The 2014-flare and optical polarization behaviour are the result of the shock exiting the 43~GHz radio core, suggesting that shock crossings are one of the possible mechanisms for EVPA rotations. 
\end{abstract}

\section{Introduction}\label{sec:intro}

The quasar 3C 454.3 is a powerful blazar at $\rm RA=22h~53m~57.7s$, Dec=+16$\rm^o~08^\prime~53.5^{\prime\prime}$, and redshift z=0.859  \citep{Hewitt1989}. It is known to be highly variable across the electromagnetic spectrum and among the brightest objects in the $\gamma$-ray sky \citep{4LAC2019}. Not surprisingly, it has been the target of multiple studies (e.g., \citealp{Bonnoli2011,Sasada2012,Jorstad2013}) since it has shown interesting and sometimes puzzling behavior. It has been classified as a Flat Spectrum Radio Quasar based on the equivalent width of its emission lines and as a low synchrotron peak source based on the location of the peak of the synchrotron emission ($\nu_{\rm peak}\sim$2.5$\times$10$^{13}$Hz \citealp{Lister2015}). It also belongs to the class of {\it rotators} as it has shown multiple rotations of the Electric Vector Position Angle (EVPA) of the optical linear polarization over the years \citep{Jorstad2010,Blinov2015,Blinov2016,Blinov2016-II,Blinov2018}. EVPA rotations are unique phenomena with so far no clear origin. Generally, they are 
considered to be the result of either random walks of the polarization vector that leads to an 
apparent coherent change of the EVPA \citep{Moore1982,Marscher2014,Kiehlmann2017} or due to deterministic processes \citep{Marscher2008,Abdo2010-III,Zhang2014,Lyutikov2017}. In this work, we investigate the nature of two prominent multi-wavelength flares in 2013 and 2014 concurrent with optical polarization variations in an attempt to better understand the origin of the multiband flaring as well as possibly shed more light on the mechanism of EVPA rotations. In section \ref{sec:obs} we describe the multi-wavelength data used in this work, in sections \ref{sec:multiwavelength} and \ref{sec:origin} we describe the events and investigate their origin and in section \ref{sec:disc} we discuss our findings. We summarize in section \ref{sum}.

\section{Observations and data reduction}\label{sec:obs}
We used the optical and near-infrared (NIR) broadband photometric data in B, V, R, I, J, H, K 
bands. The optical data where collected from six different telescopes, namely: the St. Petersburg 
University 40~cm LX-200 telescope; the Crimean observatory 70~cm AZT-8 telescope 
\citep{Larionov2008}; 1.54~m Kuiper and the 2.3~m Bok telescopes of the Steward observatory 
\citep{Smith2009}, the 1.82 m Perkins telescope (Flagstaff, AZ; VLBA-BU-BLAZAR program\footnote{\url{www.bu.edu/blazars/VLBA_GLAST/3c454.html}}); and the SMARTS monitoring program \citep{Bonning2012}. The NIR J, H, K -bands data 
were obtained at the 1.1~m telescope of the Campo Imperatore observatory \citep{Larionov2008} and the SMARTS monitoring program. All measured magnitudes have been corrected for the Galactic extinction according to \cite{Schlafly2011}.

The optical polarimetric data were taken by multiple instruments as well. The RoboPol monitoring 
program \citep{Pavlidou2014} provided R-band polarimetry (for details on the data reduction see \citealp{King2014,Panopoulou2015}). R-band polarimetry was also provided by the Perkins telescope (the data description is given in \citealp{Jorstad2010}). The St. Petersburg University monitoring program obtained ``white light'' polarimetric data at the LX-200 telescope and R-band measurements at the AZT-8 telescope. The details on their data reduction can be found in \cite{Larionov2008}. Moreover, we supplemented our dataset polarimetric data from the Steward Observatory\footnote{\url{http://james.as.arizona.edu/\~psmith/Fermi/}} \citep{Smith2009}. The values of EVPA in this work are measured from North to East following the IAU convention \citep{EVPAconv}. The 180$\rm^o$ ambiguity of EVPA was solved by minimization of the following value $|EVPA_{\rm i} - EVPA_{\rm i-1}| - \sqrt{\sigma(EVPA_{\rm i})^2 - \sigma(EVPA_{\rm i-1})^2}$ for consecutive measurements \citep[see][for related caveats]{Blinov2019}.

The $\gamma$-ray observations used in this study were taken by the Large Area Telescope (LAT) on board of the {\it Fermi} $\gamma$-ray space telescope \citep{4FGL2020}. 3C~454.3 is part of the monitored source list\footnote{\url{https://fermi.gsfc.nasa.gov/ssc/data/access/lat/msl_lc/}} and has daily and weekly binned publicly available data. For our analysis we used the daily binned light curve. 3C 454.3 is also observed as part of the LAT source monitoring program\footnote{\url{https://www.swift.psu.edu/monitoring/}} by the Neil Gehrels {\it Swift} Observatory (hereafter {\it Swift}, \citealp{Stroh2013}) in the 0.3-10 keV range. The {\it Fermi}, SMARTS, Steward observatory and {\it Swift} data are publicly available. All remaining datasets are available on demand.

\section{The multiwavelength view of 3C~454.3}\label{sec:multiwavelength}
\begin{figure*}
\resizebox{\hsize}{!}{\includegraphics[scale=1]{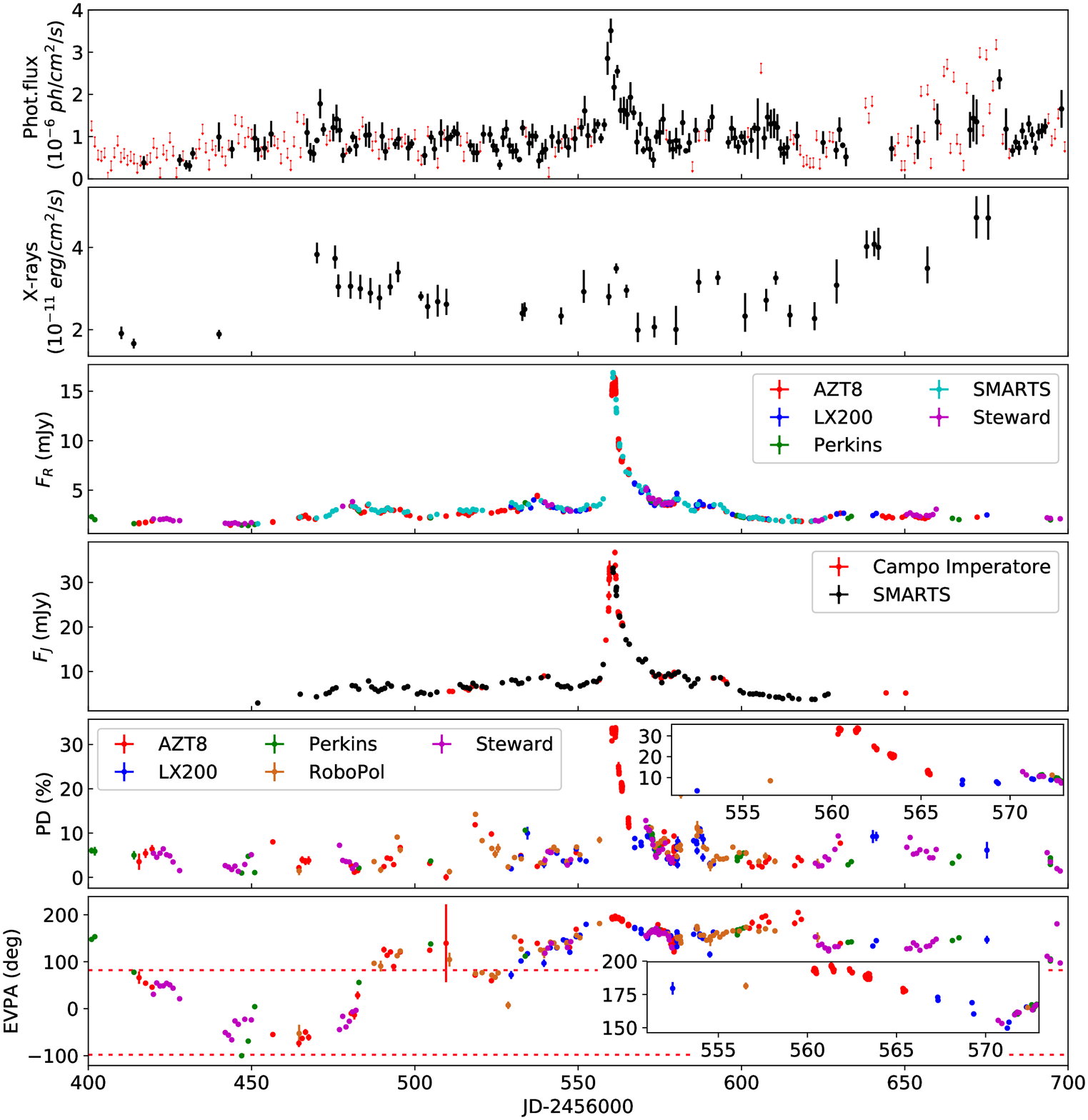} }
 \caption{Multiwalength view of the 2013 flare. Panels from the top are $\gamma$-rays, X-rays, R-band flux-density, J-band flux density, R-band degree of polarization, EVPA versus time. The dashed line in the lowest panel represents the average jet direction taking into account the 180$\rm^o$ ambiguity of the EVPA.}
\label{plt:lc_2013}
\end{figure*}

\begin{figure*}
\resizebox{\hsize}{!}{\includegraphics[scale=1]{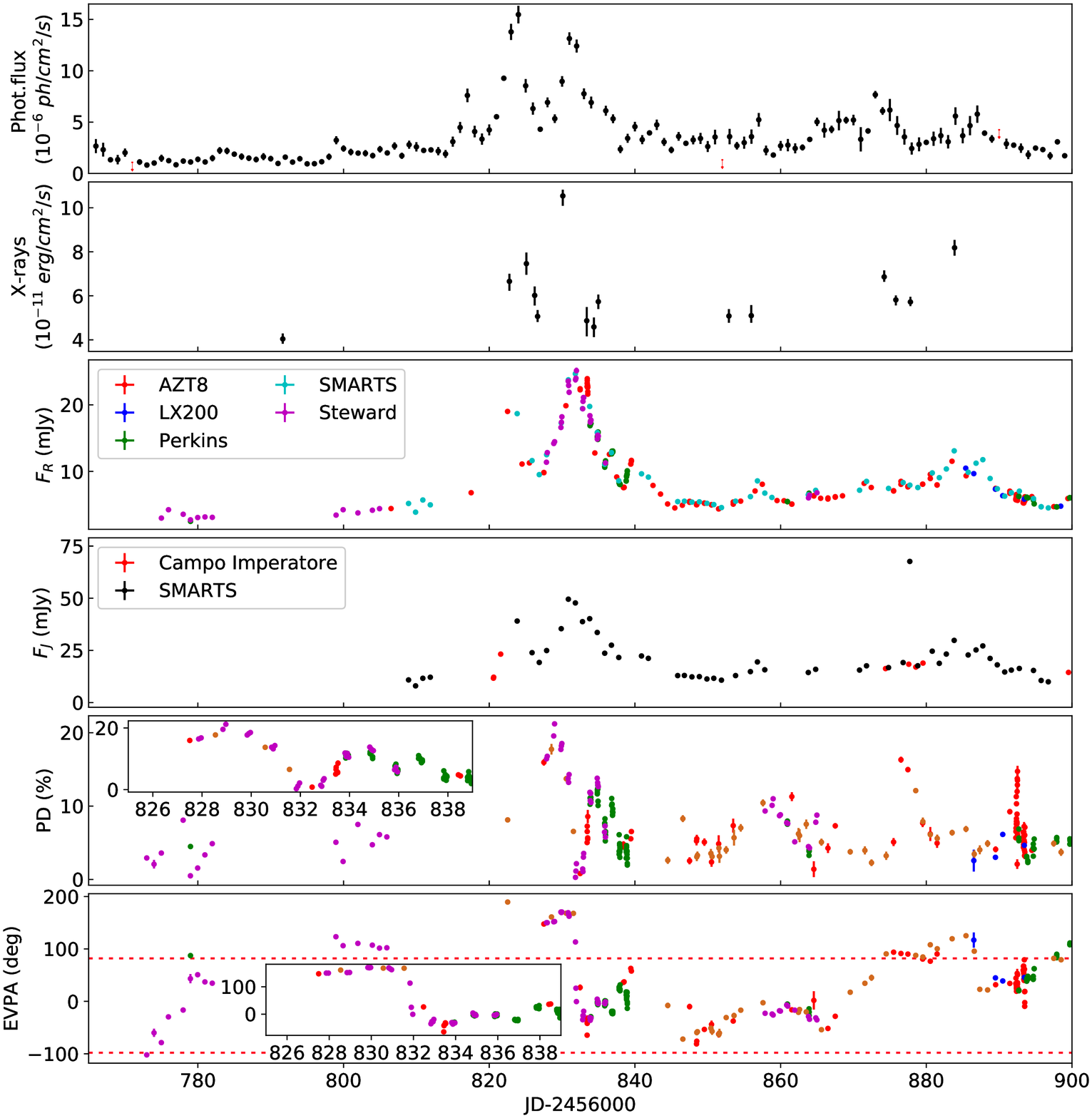} }
 \caption{Multiwalength view of the 2014 flare. Panels from the top are $\gamma$-rays, X-rays, R-band flux-density, J-band flux density,  R-band degree of polarization, EVPA versus time. The dashed line in the lowest panel represents the average jet direction taking into account the 180$\rm^o$ ambiguity of the EVPA. Color coding is the same as in Fig.\ref{plt:lc_2013}.}
\label{plt:lc_2014}
\end{figure*}

\begin{figure}
\resizebox{\hsize}{!}{\includegraphics[scale=1]{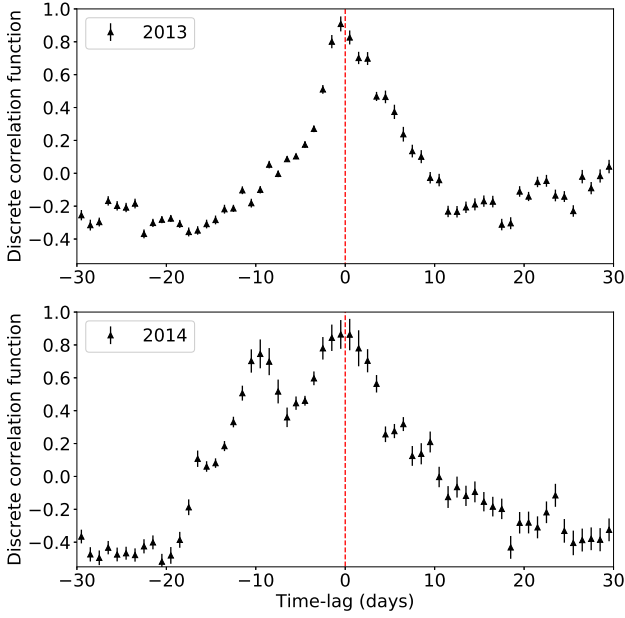} }
 \caption{Discrete correlation function between optical (R-band) and $\gamma$-rays for the 2013-(top panel) and 2014-flare (bottom panel) using a binning of 1~d}
\label{plt:dcf}
\end{figure}

\begin{figure}
\resizebox{\hsize}{!}{\includegraphics[scale=1]{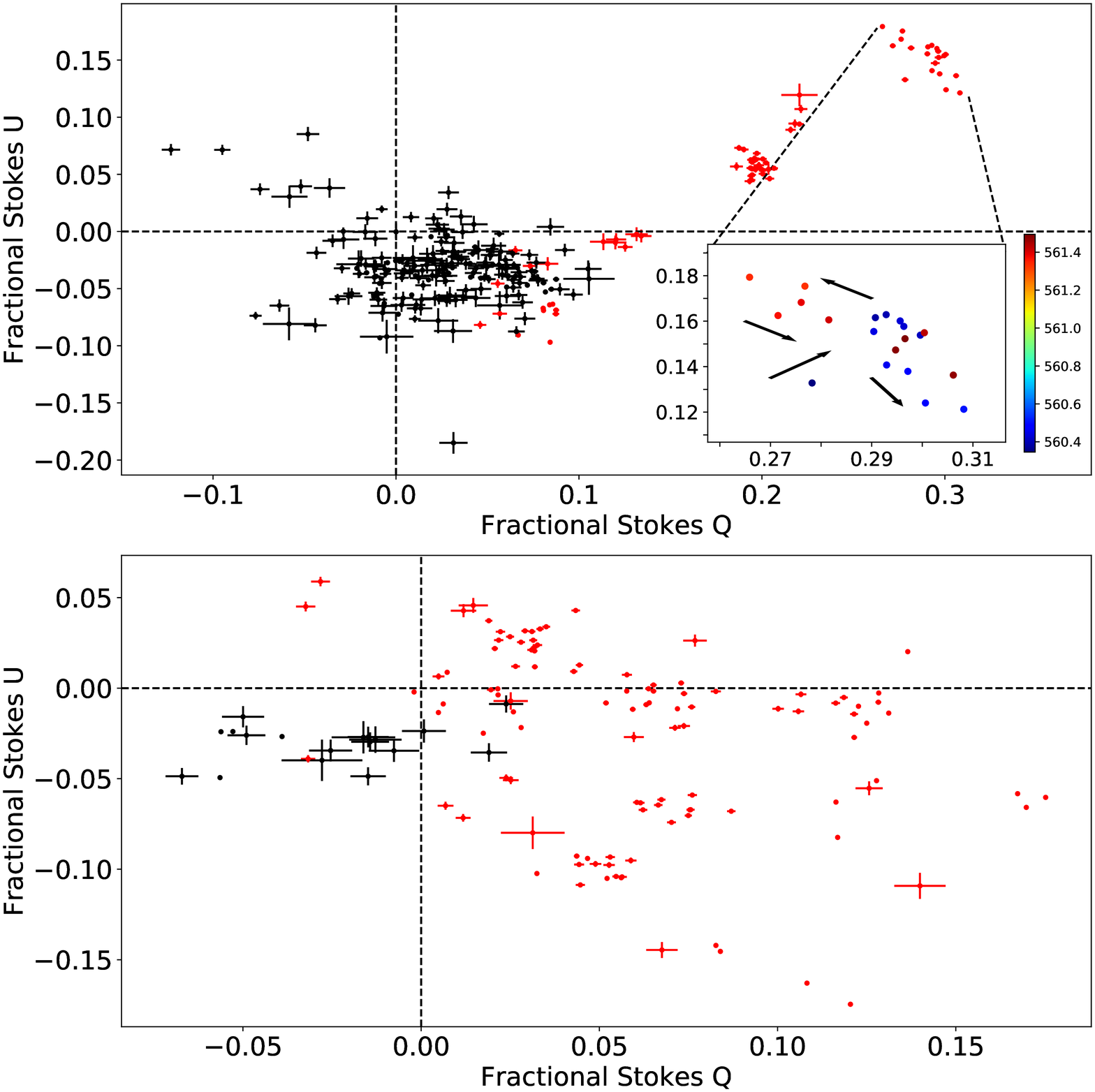} }
 \caption{Stokes Q versus Stokes U for the 2013-(top panel) and 2014-flare (bottom panel). Red is for the flaring and black for the quiescent periods. The black dashed lines mark 0-0. Inset (top panel) shows the circular motion of the Q-U vector during the flare peak. The colorbar shows the observing dates in JD-2456000.}
\label{plt:qu_plane}
\end{figure}

Figures \ref{plt:lc_2013} and \ref{plt:lc_2014} show the multiwavelength view of the 2013 and 2014 flares respectively. In both cases there are simultaneous flares in all bands as well as the polarization degree, and variations of the EVPA. The optical and $\gamma$-ray variations appear to be simultaneous. Indeed, the discrete correlation function (DCF, \citealp{Edelson1988}) in both cases yields time-lags consistent with zero (Fig. \ref{plt:dcf}). This is not surprising since 3C~454.3 shows correlated optical--$\gamma$-ray variability over long time periods ($\sim$8 years light curves, \citealp{Liodakis2018,Liodakis2019}). During the 2013-flare, the optical and $\gamma$-ray emission reaches maximum on $\sim$JD 2456560. The increase in optical flux density from the quiescent level is by an order of magnitude. The $\gamma$-ray flux increases by a factor of four. For the 2014-flare, the $\gamma$-ray maximum is at $\sim$2456823 whereas the optical maximum is at $\sim$2456830. The increase of the optical flux density from the 2013-flare is by a factor of two making it the brightest optical flare since 2008. The $\gamma$-ray flare is the fourth brightest and about a factor of three lower than the historically brightest flare in the source. There is also a visible increase in the X-ray flux during the 2013-flare, and apparent flaring during the 2014-flare  even with the much lower (compared to other bands) sampling of the X-ray light curves. The simultaneous increase is also true for the optical polarization degree for the 2013-flare reaching a maximum of $\sim34\%$ -- the highest polarization degree reached by the source since 2008; the 2014-flare shows more intriguing behavior. There are three $\gamma$-ray peaks during its duration at $\sim$2456816, $\sim$2456823, and $\sim$2456830. The optical variations, although not as well sampled, appear to be in good agreement with the $\gamma$-rays. The optical polarization maximum (21\%) lies right before the third flare ($\sim$2456829) with a sharp drop to almost zero (0.29\%) right after ($\sim$2456832). At the same time there is a roughly 233 degrees EVPA rotation. For the 2013-flare the EVPA shifts monotonically until the flare maximum when it changes direction and returns to pre-flare levels. The EVPA variations in the 2013- case does not follow the strict definition of a rotation introduced by the RoboPol program \citep{Blinov2015}. However, the 180$\rm^o$ ambiguity correction of the EVPA can be affected by the presence of multiple polarized components \citep{Ikejiri2011}. The circular motion of the Stokes Q-U vector, which is characteristic of an EVPA rotation, is then shifted away from the origin (zero-zero in the Q-U plane) that could mask the rotation in the EVPA-plane. This effect is clearly shown in Fig. \ref{plt:qu_plane} where during the flaring periods (marked with red) on both occasions the centroid is shifted from the origin. This ambiguity could also be affected by limited sampling. During the rotation in 2014 some observations are spaced by roughly 1.5 hours and we still observe $\sim80^o$ jumps during the steepest decline in the rotation. Using the Q-U plane visualization tool  {\it Timetubes\footnote{\url{https://github.com/MistletoeNaoko/TimeTubesWeb}}} \citep{Uemura2016,Fujishiro2018} we have verified that in both flaring events there is a circular motion of the Q-U vector suggesting that the observed EVPA rotations are real and not due to systematics related to e.g., sampling (an example is shown in the inset of Fig. \ref{plt:qu_plane} upper panel).

\section{Origin of the multiwavelength flares}\label{sec:origin}

\subsection{Spectral analysis}
\begin{figure}
\resizebox{\hsize}{!}{\includegraphics[scale=1]{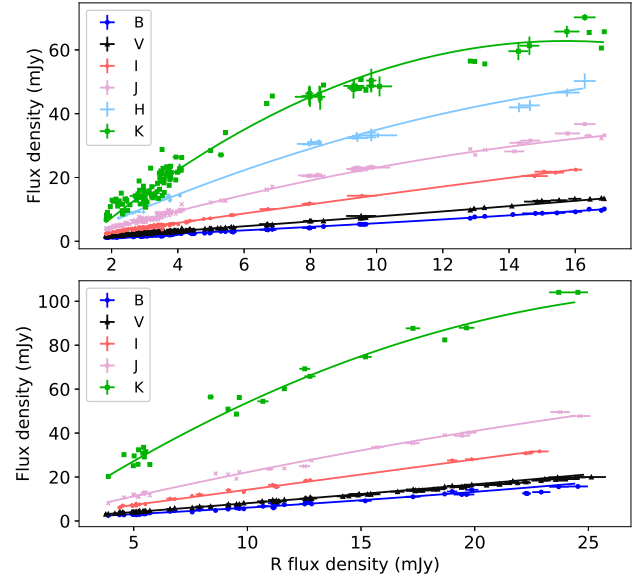} }
 \caption{BVIJHK versus R-band flux densities for the 2013- (upper panel) and 2014-flares (lower panel)}
\label{plt:flux_flux}
\end{figure}

\begin{figure}
\resizebox{\hsize}{!}{\includegraphics[scale=1]{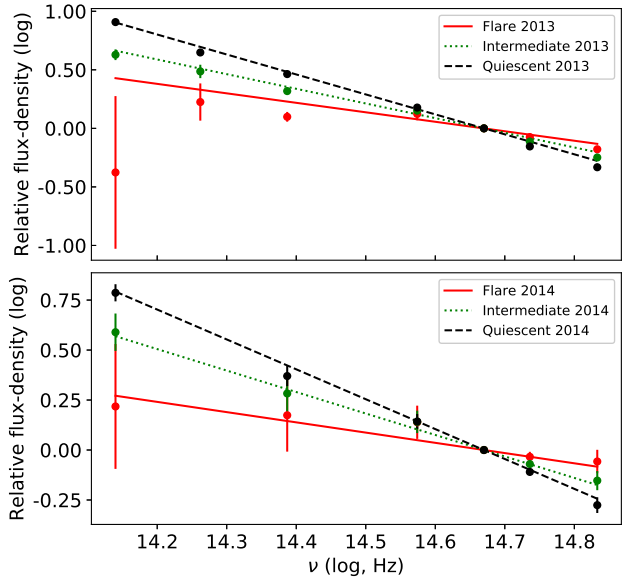} }
 \caption{Relative optical and NIR SED of the variable component for the 2013 (upper panel) and 2014 (lower panel) flares. In both panels the red (solid), green (dotted) and black (dashed) lines show flaring, intermediate and quiescent spectra.}
\label{plt:relative_spect}
\end{figure}

Multiwavelength variability concurent with EVPA rotations has been observed in a few blazars (e.g., \citealp{Marscher2008,Marscher2010}). Results by the RoboPol program suggest a strong connection between EVPA rotations and $\gamma$-ray flares \citep{Blinov2018}. From the deterministic point of view, the suggested models usually involve shocks propagating in the helical magnetic field or helical trajectories (e.g., \citealp{Marscher2008,Zhang2014,Lyutikov2017}) as well as along curved trajectories (e.g., \citealp{Abdo2010-III,Nalewajko2010}). To investigate whether shocks were responsible for the outbursts we examine their spectral behavior.  We plot the BVIJHK versus R-band flux densities for both events (Fig. \ref{plt:flux_flux}) and fit them with second degree polynomials in the form of $y=a+bx+cx^2$. We find the dependencies to be mostly linear with deviations in the infrared bands. For the 2013-flare the J- H- and K-band show significant curvature with $c=-0.064\pm0.004$, $c=-0.11\pm0.02$, and $c=-0.30\pm0.02$ respectively. For the 2014-flare only the K-band shows a hint of non-linearity at roughly $\sim3\sigma$ level with $c=-0.11\pm0.02$. The remaining colors show $c$ consistent with zero. The observed non-linearity, particular in the 2013-flare, can be interpreted as a viewing angle change of a discrete moving emission region \citep{Papadakis2007,Larionov2010,Larionov2013}. Enhanced Doppler beaming caused by a viewing angle change will increase the flux across bands while at the same time shifting the intrinsic spectrum towards higher frequencies. A pure power-law spectrum will remain unchanged; however, if the spectrum is convex (i.e. bluer-when-brighter, see Fig.~\ref{plt:relative_spect}) an increase in Doppler factor will cause the spectrum to change over time creating the observed non-linearity. Under this interpretation, following \cite{Larionov2010}, we quantify the required change in the Doppler factor to be a factor of $\sim 1.5-1.7$ for the 2013-flare and by a factor of $\sim 1.4-1.5$ for the 2014-flare depending on the jet model (continuous or discrete jet). Using the polynomial fits we also approximate the spectra for the flaring and non-flaring time intervals. We use the method described in \cite{Hagen-Thorn1994}, where the broadband
optical-to-infrared SED of the variable component is derived using flux-flux diagrams. Since in our case the flux-flux diagrams in NIR bands are curved, we used the derivatives of the polynomial fits to find the relative SED of the variable component. The result is shown in Fig.~\ref{plt:relative_spect}, where the relative SED of the variable component is plotted at three different activity states: flaring, intermediate and quiescent, where the R-band flux density was 15, 8.5 and 2 mJy for the event of 2013 and 24, 14 and 4 mJy for 2014. We find a significant hardening of the variable component SED during both events, which corresponds to bluer-when-brighter behaviour. The SED can be approximated by power law with spectral indices $-0.81\pm0.09$ and $-0.51\pm0.22$ during 2013 and 2014 flaring states. During corresponding quiescent states the spectral index was $-1.71\pm0.02$ and $-1.50\pm0.07$. We note that here we analyse only the variable component spectral changes, while in combination with underlying constant emission having different SED the color behaviour could be more complex \citep{Sasada2010} and even demonstrate the opposite redder-when-brighter behaviour \citep{Raiteri2008}.

\subsection{VLBI data analysis}

\begin{figure}
\resizebox{\hsize}{!}{\includegraphics[scale=1]{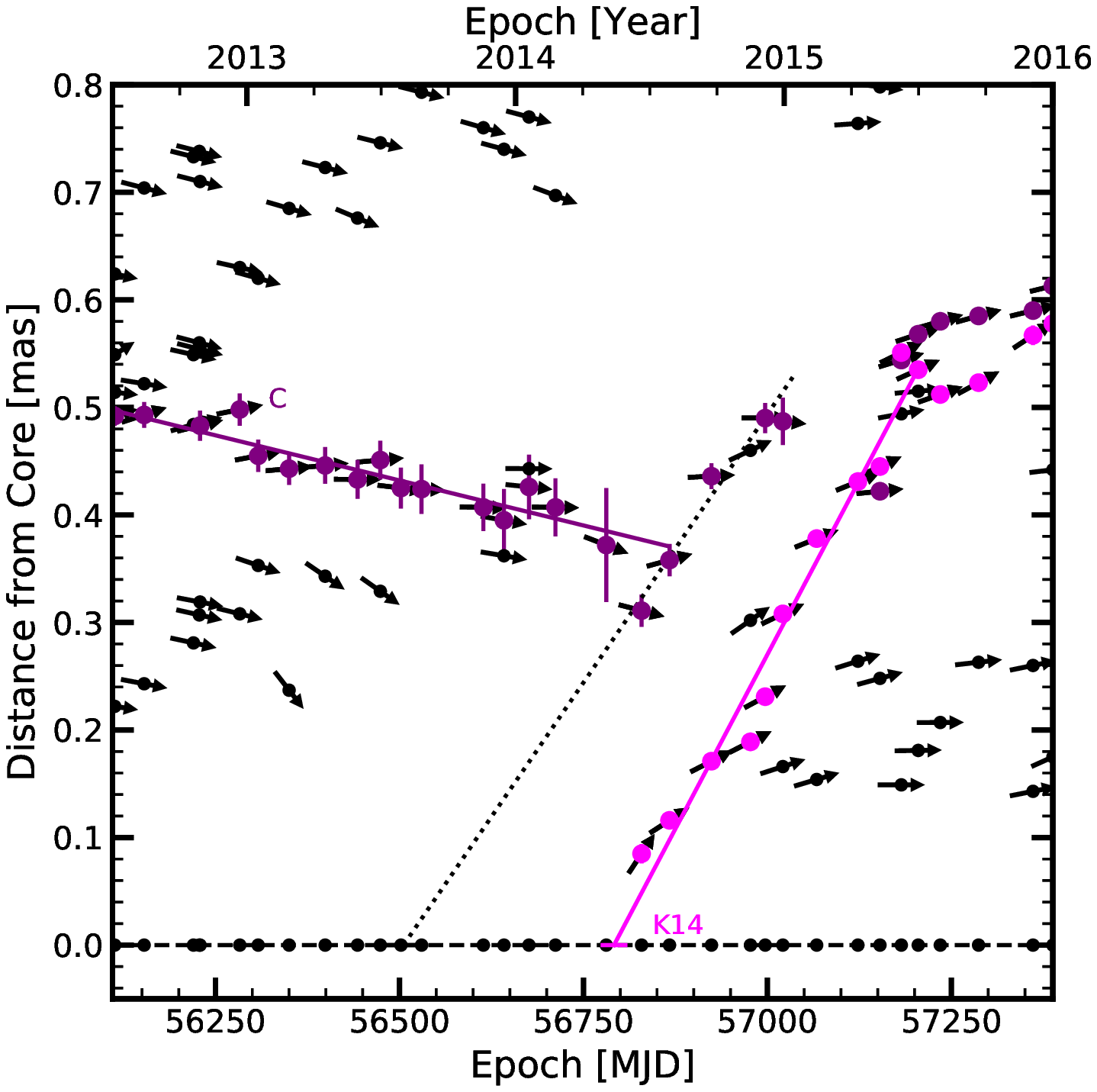} }
\resizebox{\hsize}{!}{\includegraphics[scale=1]{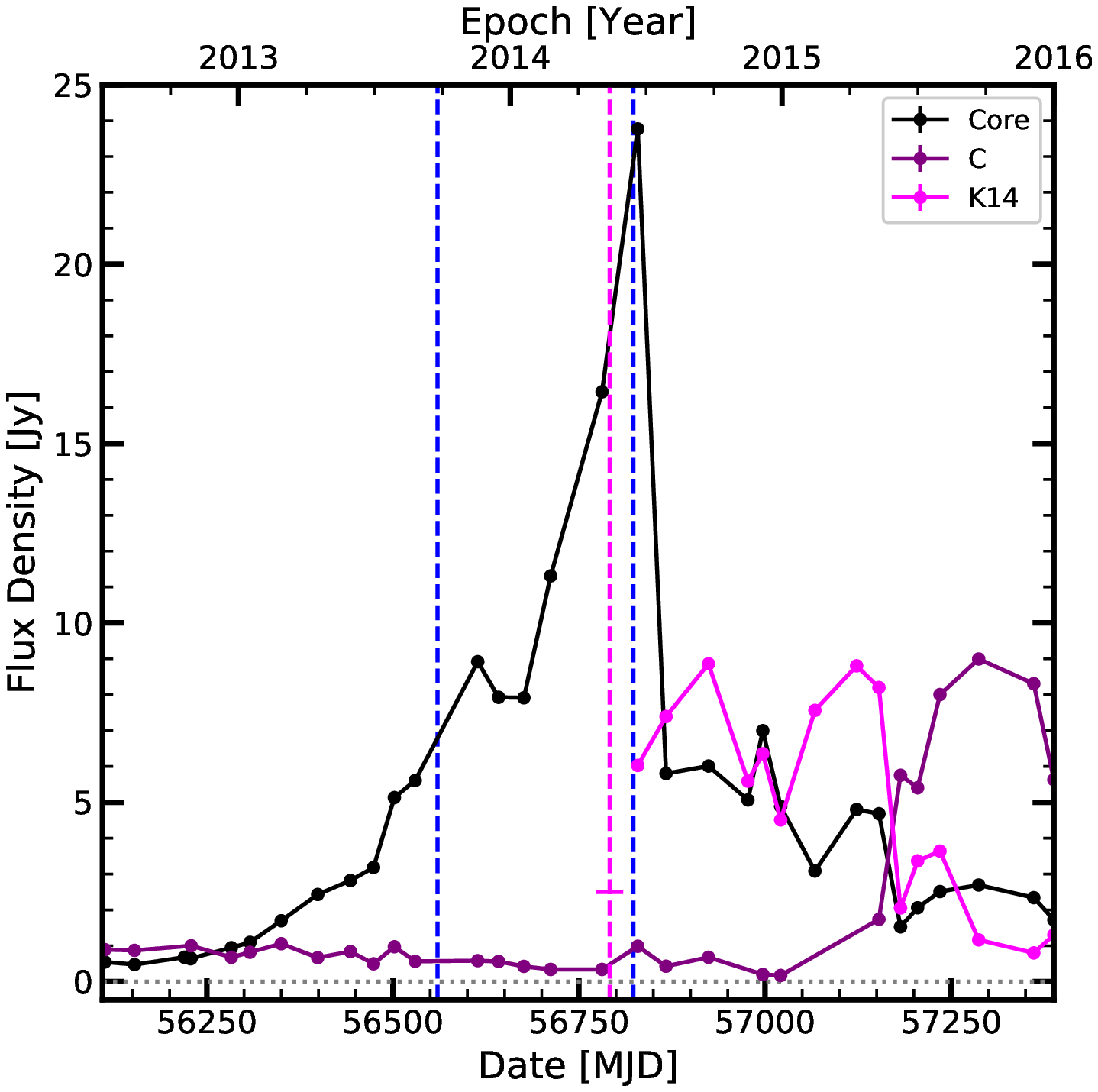} }
%\plottwo{3C454_motionN.eps}{3C454_FluxN.eps}
\caption{{\it Top:} Separation of knots C (purple circles) and K14 (pink circles) from the core A0 (black dashed line); vectors indicate position angles of knots with respect to the core;
the pink and purple solid lines approximate the motions of K14 and $C_{in}$, respectively; the dotted black line fits the motion of $C_{out}$; black circles show positions of unidentified knots. {\it Bottom:} Light curves of the core A0 (black), C (purple), and K14 (pink); the pink dashed line indicates the time of ejection of K14, while the blue dashed lines mark times of maxima of $\gamma$-ray flux during the 2013 and 2014 flares.}  
\label{move}
\end{figure}

\begin{figure}
\epsscale{0.70}
\plotone{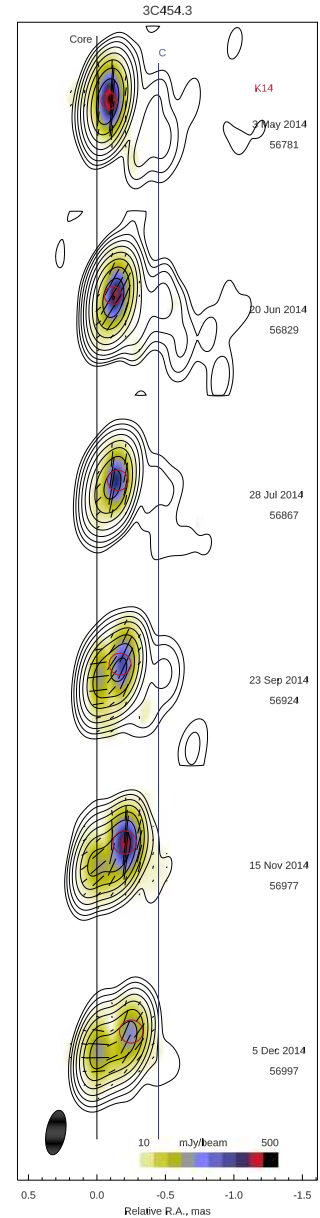}
\caption{Sequence of total (contours) and linearly polarized (color scale) intensity images of 3C454.3 at 43 GHz, convolved with a beam of FWHM dimensions 0.33$\times$0.14 mas$^2$ along PA=$-10^\circ$. The global total intensity peak is 21.7 Jy/beam and the global polarized intensity peak is 496 mJy/beam. Black line segments within each image show the position angle of polarization; the length of the segment is proportional to the polarized intensity values; the black vertical lines indicate the position of the core, A0, and quasi-stationary feature, C; the red circles mark positions of knot K14 according to model fitting.} \label{vlba14}
\end{figure}

In order to definitively confirm the presence of a shock propagating in the jet we analyze the total and polarized intensity images of the quasar 3C~454.3 derived from data obtained with the Very Long Baseline Array (VLBA) at 43~GHz  within the VLBA-BU-BLAZAR program\footnote{\url{www.bu.edu/blazars/VLBAproject.html}}. For values of the cosmological parameters H$_0$=70~km/s/Mpc, $\Omega_m$=0.3, and $\Omega_\lambda$=0.7, an angular size of 1~mas corresponds to 7.68~pc at the quasar's redshift of z=0.859. We have studied the kinematics of the parsec scale jet of the quasar over 3 yrs, from 2013 January 15 to 2015 December 5. The observations and data reduction have been performed in the same manner as in \cite{Jorstad2017}.  During the period considered here, the parsec-scale jet of the quasar possesses three main features: the core (A0), a quasi-stationary knot (C), and a superluminal knot (K14).  Parameters of the knots, obtained by fitting the data with a model consisting of circular components with Gaussian brightness distributions, are listed in Table~\ref{Kparm}. These include the average (over the time span) flux density at 43~GHz, maximum flux density, average distance and position angle with respect to the core, average angular size, proper motion, and apparent speed. The table also gives standard deviations for the average values and 1$\sigma$ uncertainty for other parameters. Uncertainties are calculated using formulas given in \cite{Jorstad2017}. Detailed models at each epoch will be presented in a study by Weaver et al. (in prep) devoted to kinematics of all sources in the VLBA-BU-BLAZAR program from 2013 to 2018.

Figure~\ref{move} (top panel) shows the motion of knots C and K14 with respect to the core, which is assumed to be a stationary feature on the eastern end of the jet. Knot C has been observed over decades: \cite{Pauliny-Toth1987} reported a feature in 10.7 GHz VLBI images located $\sim$0.6~mas from the core that did not show significant motion over 5 yrs of monitoring. Knot C was analyzed later in a number of studies \citep[e.g.,][]{Gomez1999,Jorstad2001,Jorstad2005,Jorstad2013,Jorstad2017}. It sometimes appears to change position, as seen in  Figure~\ref{move} (top panel) and reported in Table~\ref{Kparm}. Its projected distance from the core decreased by $\sim0.1$ mas over $\sim700$ days (mid-2012 to mid-2014) before it returned to $\sim0.5$ mas from the core by the end of 2014. This motion relative to the core could be connected with the motion of new superluminal knots down the jet. For example, the motion of C toward the core from 2012 to 2014 (Figure~\ref{move}, top panel) could be explained if the core region appeared to shift downstream as the highly superluminal, bright knot K14 approached the core (more slowly than it moved after
mid-2014), crossed it, and then moved downstream
while still blended with it at the resolution of the images. A contributing factor might
be an increase of the opacity of the core region, which would cause the apparent position
of the core to move downstream. The increase in flux density of the core seen in
Figure~\ref{move} (bottom panel) should have been accompanied by such an increase in opacity. According to this scenario, after K14 is resolved separately from the core, the 
core returned to its 2012 position, as did C. 
Table~\ref{Kparm} gives the parameters of C, including the proper motion for the period when it moves toward the  core, $C_{in}$, and from the core toward its stationary position, $C_{out}$. Figure~\ref{move} (bottom panel) presents light curves of the core, C, and K14. 

Knot K14 is of special interest because its time of ejection on JD: 2456797$\pm$15
agrees within 2$\sigma$ uncertainty with the maximum of the $\gamma$-ray flare on JD$\sim$2456823. In addition, the core light curve (Figure~\ref{move}, bottom panel)
shows that the core was in an elevated flux state at 43 GHz during the entire period 
from the $\gamma$-ray flare in 2013 to the $\gamma$-ray flare in 2014, and the maximum of the $\gamma$-ray flare in 2014 coincides with the maximum flux of the core. This can be interpreted as the entire $\gamma$-ray activity of these two flares being connected with 
the propagation of K14 through the VLBI core, which also implies that the 43 GHz core is a physical structure, e.g., a conical shock \citep{Marscher2016}, and K14 is a moving shock. Superluminal knots in the jet of 3C454.3 have a range of speeds, with proper motions from 0.14 to 0.53~mas/yr \citep[e.g.,][]{Jorstad2001}. Therefore, knot K14, with 0.47~mas/yr, is
one of the fastest superluminal features observed in the jet, although it decelerates significantly as it approaches C. Figure~\ref{vlba14} shows a sequence of VLBA images of the 3C454.3 jet, which exhibits the motion of K14 from the core toward C.

Figure~\ref{vlba14} also reveals that K14 is the most polarized feature in the jet, with EVPA perpendicular to the jet direction, which implies that the magnetic field is parallel to the jet axis. Such a direction of magnetic field is common for superluminal knots observed between A0 and C in 3C454.3 \citep{Jorstad2013}. However, such a magnetic field direction poses a problem for the common shock-in-jet model \citep[e.g.,][]{Hughes1985,Marscher1985} to explain the nature of K14. Since a shock compresses the magnetic field component that is transverse to 
the shock normal, the projected direction of the field should be perpendicular to the
jet axis, so that the EVPA is parallel to the jet. In order for the field to be parallel
to the jet, the magnetic field of the ambient jet plasma would need to be well 
ordered in that direction prior to the passage of the shock. The compression of any
disordered component of the field by the shock could then be insufficient to overcome the parallel field. In this case, however, the increase in flux density should be relatively
modest, since it would not include the effects of a major increase in field strength. The
high flux of K14 would then be caused solely by the compression of the relativistic
electron density and higher bulk Lorentz factor of the knot. The $\gamma$-ray flux from
external Compton scattering is sensitive to these two factors, but not to the magnetic
field strength \citep{Sikora2009}, hence the strong $\gamma$-ray outburst can be 
understood under this scenario.

Knowing the proper motion of K14 within the core region, we can estimate its entrance ($T_{in}$) and exit ($T_{out}$) time from the core using the sizes of K14 and A0 (Table~\ref{Kparm}) and taking into account that the ejection time of K14 ($T_\circ$) corresponds to the passage of the centroid of K14 through the centroid of A0. We consider three different proper motions for K14 to characterize its motion within the core region: 1) $\mu$ of K14 assuming that the speed of K14 is the same within the core as detected later in the jet, 2) $\mu$ of C$_{in}$ assuming that motion of $C$ toward the core reflects the motion of K14 as it approaches the core, and 3) $\mu$ of C$_{out}$ assuming that the motion of $C$ back to the stationary position can be connected to the motion of K14 inside the core. The derived values of $T_{in}$ and $T_{out}$ are listed in Table~\ref{Tcalc}. Analysis of Tables~\ref{Kparm} \& \ref{Tcalc} shows that the $\gamma$-ray flare in 2014 can be connected with the exit of knot K14 from the core. We can estimate the distance traveled by K14 ($\rm \Delta{r_{K14}}$) between the 2013- and 2014-flares using,
\begin{equation}
\Delta{r_{K14}}=\frac{\beta_{app}c\Delta{t}}{(1+z)\sin\theta},
\end{equation}
where $\rm \beta_{app}$ is the apparent velocity, $\Delta{t}$ is the time between the 2013-flare and the first peak of the 2014-flare, $\theta$ is the viewing angle. \cite{Jorstad2010} and \cite{Pushkarev2012} provide estimates of the distance between 
the 43/15~GHz VLBI core and the BH, $\sim$18~pc and $\sim$20~pc, respectively. 
This puts a limit on ${\Delta}r_{K14}<$18~pc, if we associate the 2013-flare with the propagation of K14 in the jet. Using $\beta_{app,Cin}=6.5$ (Table \ref{Kparm}) and $\theta\sim2.6$ as found by \cite{Weaver2019}, which agrees with the maximum viewing angle given by \cite{Liodakis2018-II}, we have derived a distance between the locations of the 2013 and 2014 events corresponding to $\sim$16~pc. The later places the dissipation zone during the 2013-flare within $\sim$2~pc from the BH.  

\section{Interpretation \& Discussion}\label{sec:disc}
\begin{figure}
\resizebox{\hsize}{!}{\includegraphics[scale=1]{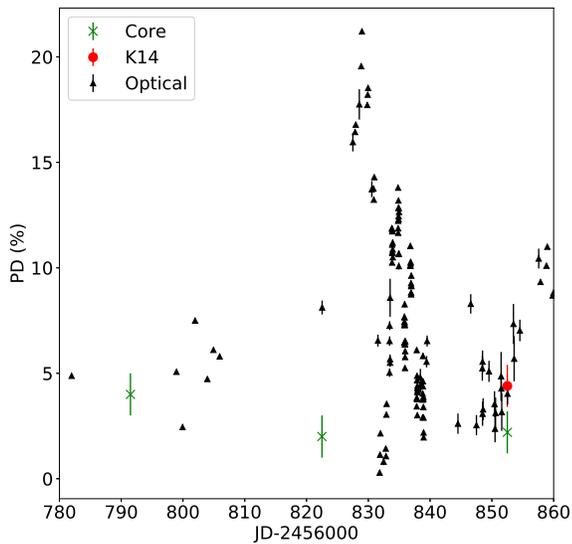} }
 \caption{Radio core (green ''x''), K14 (red ``$\bullet$''), and optical (black ``$\triangle$'') polarization degree for the 2014-flare.}
\label{plt:radio_opt_pol}
\end{figure}
\begin{figure}
\resizebox{\hsize}{!}{\includegraphics[scale=1]{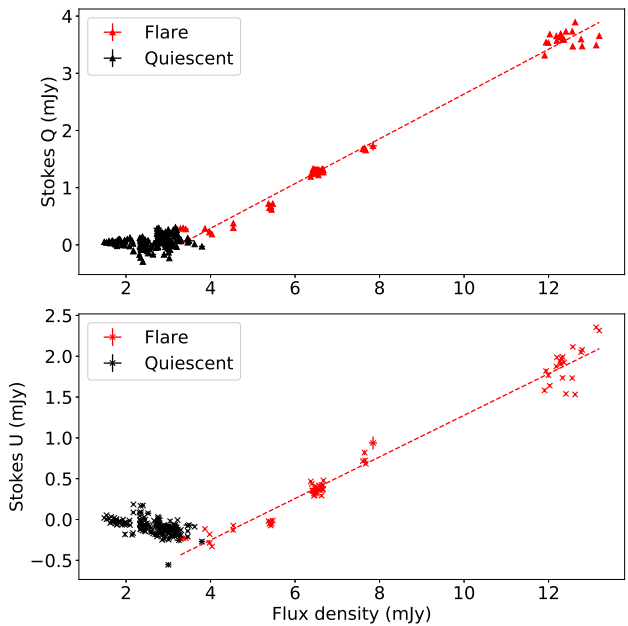} }
 \caption{Stokes Q (top panel) and Stokes U (bottom panel) versus total intensity for the 2013-flare. Black is for observations during quiescence and red for observations during the flaring event. The red dashed line shows the best-fit relation.}
\label{plt:QU_vs_I}
\end{figure}

The results of the above spectral and VLBI analysis demonstrate that a shock was propagating in the jet of 3C 454.3 causing the multiband and polarization flaring as well as EVPA variations in two separate occasions before finally being ejected from the 43~GHz radio core. 

The origin of the 2014-flare can be interpreted as the result of the travelling shock crossing a standing recollimation shock that is typically observed as the radio core \citep{Daly1988}. This is evident by the ejection of the radio component K14 as well as the hardening of the spectrum. The drastic drop in the optical polarization degree and the concurrent EVPA rotation could be either due the shock moving in a helical path \citep{Larionov2013-II} or due to the change in emission dominance between two polarized components (i.e., standing and moving shock) with orthogonal magnetic field orientations \citep{Cohen2020}. This would also suggest that shock-crossing is one of the mechanisms responsible for EVPA rotations. The three-peak profile of the 2014-flare is also most likely due to the shock-shock interaction. Such sub-flaring has been noted before in 3C~454.3 \citep[e.g.,][]{Jorstad2010, Weaver2019} as well as other blazars (e.g., \citealp{Larionov2013-II,MAGIC2018}) suggesting that it is a common phenomenon of blazar jets. Figure \ref{plt:radio_opt_pol} shows the core and K14 polarization degree contrasted against the optical. The polarization orientation of the core and K14 remained roughly perpendicular to the average jet direction before and after the event while the pre- and post-flare polarization levels are consistent between radio and optical within the uncertainties. This would suggest that the extreme behavior seen in the optical polarization is most likely due to changes in the uniformity of the magnetic field further supporting the shock-shock interaction interpretation. 

The 2013-flare is more difficult to interpret. From Fig. \ref{plt:lc_2013} the source was in a quiescent state for an extended period of time, has a single broadband flare and then returned to quiescence before the shock exited the core producing the 2014 event. Assuming the distance between the black hole and the radio core is $\rm 10^5 R_g$ \citep{Marscher2008}, where $\rm R_g$ is the gravitational radius, for a $10^{9.34}M_{\odot}$ black hole \citep{Liodakis2017-III,Liodakis2020} the 2013-flare would occur $\rm \sim 3\times10^3R_g$ downstream from the black hole. This would place the flaring region at the edge or beyond the broad-line region (BLR) and at the acceleration and collimation zone \citep{Marscher2008,Marscher2010}. One possible interpretation would be that the 2013-flare was the result of the shock formation. In this case, given the fact that the apparent displacement of knot C had begun in 2012 would suggest the existence of propagating plasma cells that merged creating the 2013-flare (e.g., \citealp{Spada2001}). Partial dissipation through magnetic reconnection of accelerating stripes \citep{Giannios2019} could also produce a smaller flare (2013-flare) before the main dissipation event (in this case the 2014-flare). However, the observed polarization flare is not consistent with the polarization expectations from magnetic reconnection events \citep{Zhang2018}.

To further investigate the origin of the 2013-flare we use the simultaneous (within one day) optical and $\gamma$-ray observations to derive a scaling between the fluxes in log-log space. We find $F_\gamma\propto F_{opt}^{1.04\pm0.37}$, suggesting a linear scaling. In a simple scenario where the $\gamma$-ray emission would be the result of particle injection in the jet under constant magnetization the linear scaling would point to external Compton emission. Possible sources of external photons could be either from the BLR or the dusty torus, or a localized source (ring of fire model, \citealp{MacDonald2015}). In the latter case we do not expect a significant increase of the optical flux contrary to the observed behavior making this scenario unlikely. In the former case, the observed flare in optical polarization would require some change in the magnetic field configuration which would be consistent with shock formation as discussed above. Kink instabilities in the jet can also drive flux and polarization variations similar to shocks \citep{Nalewajko2017,Zhang2017}. However, such models typically produce a drop in polarization during rotation events which although consistent with the behavior of the 2014-flare, the aforementioned flare in polarization during the 2013-flare as well as the overall VLBI behavior make the propagating disturbance scenario more likely. Alternatively, the linear variations can be understood by changes in the viewing angle of the shock propagating along a bent jet or helical trajectory. In this case, the resulting variations in the Doppler factor would cause simultaneous linear fluctuations across bands \citep{Larionov2016} and can also explain the observed EVPA behavior (e.g., \citealp{Raiteri2017,Lyutikov2017,Uemura2017}). This is further supported by the fact that the 2013-flare showed significant non-linear flux-density dependencies in the infrared bands (as opposed to the 2014-flare). In the case of a supersonic flow,  an oblique standing shock would decrease the component of the flow velocity parallel to the shock normal causing the flow to bend. This would lead to compression of the magnetic field and particle acceleration (in addition to a Doppler factor change) creating conditions favorable for an outburst similar to the one observed. After the bend, rarefaction should lead
to a more quiescent state consistent with the observed behavior.

Under the shock-in-jet model we can use the R-band and polarization degree light curves to constrain the change in the beaming properties of the shock  \citep{Hughes1991,Hagen-Thorn2008}.  If the changes of the Doppler factor are due to changes in the viewing angle with respect to the shocks path in the jet, then the intrinsic viewing angle of the 
shock (taking into account aberration effects) is given by, 
\begin{equation}
\Psi=\arctan\left(\sin\theta/[\Gamma(\cos\theta)-\sqrt{1-\Gamma^{-2}]}\right),
\end{equation}
where $\theta$ is the viewing angle. The observed polarization degree would be,
\begin{equation}
p=\frac{a+1}{a+5/3}\frac{(1-\eta^{-2})\sin^2\Psi}{2-(1-\eta^{-2})\sin^2\Psi},
\end{equation}
where $\eta$ is the density ratio of shocked and unshocked regions. Assuming that the observed flux due to the shock is $F=F_o\nu^{-a}\delta'^{(2+a)}\delta^{(3+a)},$ where $\nu$ is the frequency, $F_o=F_{max}\nu^a/\delta_{max}^{3+a}$ where $\delta_{max}$ is the 
maximum Doppler factor at the peak of the flare, and $\delta'\approx 1$ is the  Doppler factor of 
the shocked plasma in the rest-frame of the shock \citep{Hagen-Thorn2008}. $\delta_{max}$ is 
estimated assuming in each case the maximum $\eta$ that produces the maximum observed polarization 
for $\psi=90^o$.  Using the above equations, we can derive an estimate of $\delta$ as a function of 
time for the duration of the flare. Assuming $\beta_{app}=6.5$ and $\theta=2.6$ degrees (which translates to $\delta_{max}=15.7$ and $\Gamma=9.2$) we estimate a change of the Doppler factor by a factor of $\sim1.5$ between the flaring and non-flaring states. A similar result was found in \cite{Sasada2012} for an outburst in 2009 ($\sim 1.4$) when assuming that all of the enhanced emission is due to changes in the Doppler factor, further supporting the change-in-Doppler-factor interpretation. Assuming constant bulk velocity that would suggest a change in the viewing angle by a factor of $\sim 2$ ($\theta_{max}\approx1.3$). Changes in the bulk Lorentz factor could also be possible, however, based on the observed behavior it would suggest that the shock experienced a sudden boost of acceleration and then immediately decelerated to its original velocity which is unlikely.

We can also constrain the polarization properties of the shock by assuming the observed emission is the superposition of two components, one constant (from the underlying jet emission) and one variable (the travelling shock). If the polarization properties of the shock remain constant, following \cite{Hagen-Thorn1999} we can write the Stokes parameters as $Q=p_{Q,fl}I+(Q_{n-fl} - p_{Q,fl}I_{n-fl})$, and  $U=p_{U,fl}I+(U_{n-fl} - p_{U,fl}I_{n-fl})$, where $I$ is the total intensity and $ I_{n-fl}$ is the intensity of the non-variable component. In this simple model we expect a linear relation in the Q-I and U-I planes with the slopes of the best-fit relation equal to the polarization fraction of the variable component (Fig. \ref{plt:QU_vs_I}). We find $p_{Q,fl}=0.255\pm0.005$  and $p_{U,fl}=0.391\pm0.004$, hence the polarization of the variable component is constrained to $p_{var}=47\%$.

\section{Summary}\label{sum}

Our results point to a single shock propagating in the jet of 3C~454.3 causing flares in multiple bands and polarization variations in 2013 and 2014. The first multiwavelength flaring event occurred as the shock propagated in the blazar's acceleration zone producing simultaneous flaring across band including the optical polarization degree. The event was most likely due to a change in the viewing angle of the shock with respect to the observer. The change of the viewing angle could be attributed to an intrinsic bend, a twisted inhomogeneous jet \citep{Raiteri2017}, a jet carrying a helical magnetic field with a variable direction/velocity \citep{Lyutikov2017}, or due to a mini-jet in a striped jet model with moderate magnetization \citep{Giannios2009,Giannios2019}. We constrain the change of $\delta$ to be by a factor of $\sim1.5$. It then continued its journey until 2014 when it caused a second flaring event while exiting the core as revealed by the 43~GHz VLBA imaging. Exiting the core was accompanied with a rotation of the optical polarization plane and strong optical polarization degree variations from $\sim 0-20\%$. The optical polarization behaviour points to two interacting components, presumably shock-shock interaction, as one of the underlying mechanisms for EVPA rotations. While two component models predict $n\times\pi$ rotations, it requires orthogonal EVPA orientations. Smaller or larger amplitude rotations are possible if the EVPAs are not orthogonal. The $\sim230^o$ rotation discussed in this work lies within the theoretical range of possible rotation amplitudes \citep{Cohen2020}. Additional factors contributing to the departure from $\sim180^o$ could be measurement uncertainties, intrinsic variability, or significant contribution from an underlying multizone turbulent jet \citep{Marscher2014,Peirson2018,Peirson2019}. Alternatively, the merger of reconnection plasmoids in a moving stripe could reproduce $>180^o$ rotations with similar polarization behavior \citep{Zhang2018}. Further  detailed modeling of the rotations and changes to the broadband emission is necessary to further pin down the intricacies of the jet emission and magnetic field geometry.

\acknowledgements
We thank the anonymous referee for useful comments that helped improve this work. I.L. would like to thank the University for Hiroshima for their hospitality during which parts of this work were completed. RoboPol is a collaboration involving the University of Crete, the Foundation for Research and Technology – Hellas, the California Institute of Technology, the Max-Planck Institute for Radioastronomy, the Nicolaus Copernicus University, and the Inter-University Centre for Astronomy and Astrophysics. This paper has made use of up-to-date SMARTS optical/near-infrared light curves that are available at \url{www.astro.yale.edu/smarts/glast/home.php}. Data from the Steward Observatory spectropolarimetric monitoring project were used. This program is supported by Fermi Guest Investigator grants NNX08AW56G, NNX09AU10G, NNX12AO93G, and NNX15AU81G. The \textit{Fermi}-LAT Collaboration acknowledges support for LAT development, operation and data analysis from NASA and DOE (United States), CEA/Irfu and IN2P3/CNRS (France), ASI and INFN (Italy), MEXT, KEK, and JAXA (Japan), and the K.A.~Wallenberg Foundation, the Swedish Research Council and the National Space Board (Sweden). Science analysis support in the operations phase from INAF (Italy) and CNES (France) is also gratefully acknowledged. This work performed in part under DOE Contract DE-AC02-76SF00515. D.~B. and S.~K. acknowledge  support  from  the  European  Research  Council  under  the  European  Union’s  Horizon  2020  research  and  innovation  program,  under  grant agreement No~771282. The research at BU is partly supported by Fermi GI grants 80NSSC17K0649 and 80NSSC19K1505. The VLBA is an instrument of the National Radio Astronomy Observatory. The National Radio Astronomy Observatory is a facility of the National Science Foundation operated by Associated Universities, Inc. This research has made use of the NASA/IPAC Extragalactic Database (NED), which is operated by the Jet Propulsion Laboratory, California Institute of Technology, under contract with the National Aeronautics and Space Administration.

\facilities{Campo Imperatore Observatory, Crimean Observatory,  {\it Fermi}, LX-200, Perkins telescope RoboPol, SMARTS, Steward Observatory, {\it Swift}, VLBA}.

\bibliographystyle{aasjournal}
% Use the LaTeX power, use bibtex properly.
%\bibliography{bibliography} %graphy.bib}%,bibliography_export.bib}

\appendix

\section{VLBI analysis results}
Table \ref{Kparm} and \ref{Tcalc} summarize the results of the  VLBI analysis of the 43~GHz images obtained within the VLBA-BU-BLAZAR program.

\begin{deluxetable}{lrrrrr}
\singlespace
\tablecolumns{6}
\tablecaption{\bf Parameters of Knots  \label{Kparm}}
\tabletypesize{\footnotesize}
\tablehead{
\colhead{Parameter}&\colhead{$A0$}&\colhead{$C$}&\colhead{$C_{in}$}&\colhead{$C_{out}$}&\colhead{$K14$}}
\startdata
$N$&26&23&13&11&10 \\
$\langle S\rangle$, Jy&5.80$\pm$5.02&2.10$\pm$2.90&0.67$\pm$0.25&3.70$\pm$3.61&6.54$\pm$2.12 \\
$S_{\rm max}$, Jy&23.77$\pm$0.45&9.0$\pm$0.83&1.06$\pm$0.22&9.0$\pm$0.83&8.86$\pm$0.51 \\
$\langle R\rangle$, mas&0.0&0.455$\pm$0.075&0.415$\pm$0.039&0.49$\pm$0.10&0.29$\pm$0.16 \\
$\langle\Theta\rangle$, deg&\nodata&$-$87$\pm$11&$-$93$\pm$9&$-$83$\pm$11&$-$61$\pm$10 \\
$\langle a\rangle$, mas&0.076$\pm$0.041&0.24$\pm$0.11&0.32$\pm$0.06&0.15$\pm$0.07&0.10$\pm$0.02 \\
$\mu$,mas~yr$^{-1}$&\nodata&\nodata&0.138$\pm$0.009&0.209$\pm$0.005&0.471$\pm$0.003 \\
$\beta_{\rm app}$, $c$&\nodata&\nodata&6.5$\pm$0.4&9.8$\pm$0.2&22.05$\pm$0.14 \\
$T_\circ$, JD &\nodata&\nodata&\nodata&2456500$\pm$42&2456797$\pm$15 \\
\enddata
\tablecomments{$N$ - number of epochs at which component was detected; for K14 $N$ corresponds to epochs of ballistic motion;
$\langle S\rangle$ - average flux density and its standard deviation; $S_{\rm max}$ - maximum flux density and its 1$\sigma$ uncertainty; $\langle R\rangle$ - average distance from the core and its standard deviation; $\langle\Theta\rangle$ - average position angle of component with respect to the core in projection on the plane of the sky and its standard deviation;
$\langle a\rangle$ - average angular size of component and its standard deviation;
$\mu$ - proper motion and its 1$\sigma$ uncertainty; $\beta_{\rm app}$ - apparent speed and its 1$\sigma$ uncertainty; $T_\circ$ - time of ejection and its 1$\sigma$ uncertainty.}
\end{deluxetable}

\begin{deluxetable}{lrrr}
\singlespace
\tablecolumns{4}
\tablecaption{\bf Epochs of K14 entrance and exit in/from the core \label{Tcalc}}
\tabletypesize{\footnotesize}
\tablehead{
\colhead{Time}&\colhead{$\mu_{K14}$}&\colhead{$\mu_{Cin}$}&\colhead{$\mu_{Cout}$}}
\startdata
$T_{in}$, JD&2456730$\pm$24&2456570$\pm$82&2456647$\pm$54\\
$T_{out}$, JD&2456864$\pm$24&2457024$\pm$82&2456947$\pm$54\\
\enddata
\tablecomments{$T_{in}=T_\circ-\Delta T$ and $T_{out}=T_\circ+\Delta T$,
where T$_\circ$ is the ``ejection'' time when the brightness centroid of K14 crossed that
of the core, and
$\Delta T \equiv (\langle a_{K14}\rangle+\langle a_{A0}\rangle)/2/\mu$. The
values of $\langle a\rangle$ are given in Table \ref{Kparm}.}
\end{deluxetable}

\end{document}